\documentclass[11pt]
{elsarticle}

\usepackage{psfrag}
\usepackage{graphicx}
\usepackage{array}
\usepackage{url}
\usepackage{amsmath,amsfonts}
\usepackage{color}
\usepackage{fullpage}

\newcommand\bp{{\mathbf p}}

\newcommand\bx{{\mathbf x}}

\newcommand\bu{{\mathbf u}}
\newcommand\bnabla{{\boldsymbol \nabla}}
\newcommand\pnabla{\bnabla_\bp}
\newcommand\tr{\mathop{\text{tr}}}

\newcommand\e{{\mathbf e}}

\newtheorem{thm}{Theorem}

\newcommand{\proof}[1]{\medskip\noindent{\em #1}}
\newcommand{\bfeqref}[1]{(\ref{#1})}

\begin{document}

\title{The Fast Exact Closure for Jeffery's Equation with Diffusion}

\author[mu-ma]{Stephen Montgomery-Smith}
\author[baylor]{David Jack\footnote{Corresponding Author.  Email: {\ttfamily david\_jack@baylor.edu}.  Phone: 254-710-3347.}}
\author[mu-mae]{Douglas E. Smith}
\address[mu-ma]{Department of Mathematics, University of Missouri, Columbia MO 65211, U.S.A.}
\address[baylor]{Department of Mechanical Engineering, Baylor University, Waco TX 76798, U.S.A.}
\address[mu-mae]{Department of Mechanical and Aerospace Engineering, University of Missouri, Columbia MO 65211, U.S.A.}

\begin{abstract}
Jeffery's equation with diffusion is widely used to predict the motion of concentrated fiber suspensions in flows with low Reynold's numbers. Unfortunately, the evaluation of the fiber orientation distribution can require excessive computation, which is often avoided by solving the related second order moment tensor equation.  This approach requires a `closure' that approximates the distribution function's fourth order moment tensor from its second order moment tensor.  This paper presents the \emph{Fast Exact Closure} (FEC) which uses conversion tensors to obtain a pair of related ordinary differential equations; avoiding approximations of the higher order moment tensors altogether.  The FEC is exact in that when there are no fiber interactions, it exactly solves Jeffery's equation.  Numerical examples for dense fiber suspensions are provided with both a Folgar-Tucker (1984) diffusion term and the recent anisotropic rotary diffusion term proposed by Phelps and Tucker (2009). Computations demonstrate that the FEC exhibits improved accuracy with computational speeds equivalent to or better than existing closure approximations.

\end{abstract}


\maketitle

\thispagestyle{empty}

\section{Introduction}
The industrial demand has continued to increase for high-strength, low-weight, rapid production parts such as those made of short discontinuous fiber composites with injection molding processes.  For effective design, it is essential to understand the dependance of the final part performance of short-fiber injection molded composites with the variations in the microstructure due to the processing (see e.g. \cite{verweyst:99,fan:98}).  The Folgar and Tucker model of isotropic diffusion \cite{folgar:84} for fiber interactions within a suspension has been used for several decades to compute fiber orientation and has been implemented to some extent within most related industrial and research computer simulations. Unfortunately, direct computations of the isotropic diffusion model are computationally prohibitive, and most implementations employ the orientation tensor approach of Advani and Tucker \cite{advani:87a} where the moments of the fiber orientation are solved, thus indirectly quantifying the fiber orientation distribution. The orientation tensor approach requires knowledge of the next higher-order moment tensor, thus requiring some form of a closure. The hybrid closure of Advani and Tucker \cite{advani:87a} has been used extensively due to its computational efficiencies, but in implementation it will overpredict the alignment state in simple shear flow \cite{jack:07a}. Cintra and Tucker \cite{cintra:95} introduced the class of the orthotropic closures, which result in significant accuracy improvements when compared to the hybrid closure, but at an increase in computational costs.

With recent advances in part repeatability, the limitation of the isotropic diffusion model has become apparent \cite{tucker:04}. Recent anisotropic diffusion models \cite{wang:08,phelps:09,koch:95,jack:08c} propose new forms with greater accuracies for modeling fiber collisions, but these anisotropic diffusion models pose a new set of computational complications. In particular is the concern that nearly all of the fitted orthotropic closures are obtained by fitting orientation information based on direct numerical solutions of the Folgar-Tucker diffusion model.  The exception is the orthotropic closures of Wetzel \cite{wetzel:99b} and VerWeyst \cite{verweyst:98} which were both constructed on distributions formed through the elliptic integral form for orientations encompassing the eigenspace \cite{cintra:95}.

The Exact Closure of Montgomery-Smith {\it et al.} \cite{montgomery-smith:10b} presents an alternative to the classical closure form, and provides an exact solution for pure Jeffery's motion (i.e., the dilute regime). The Exact Closure avoids the curve fitting process required to define fitted closures, by solving a set of related ODEs of the fiber orientation. In the present paper, we extend the Exact Closure form to systems of concentrated suspensions that are more relevant to modeling the processing of short-fiber composites.  Furthermore, we introduce the new \emph{Fast Exact Closure} (FEC) that defines conversion tensors that lead to a coupled system of ordinary differential equations that avoid costly closure computations.  The FEC form is derived for fiber collision models for both the isotropic diffusion model of Folgar and Tucker and the recent anisotropic diffusion model of Phelps and Tucker \cite{phelps:09}. Results presented will demonstrate the effectiveness of this alternative approach for modeling fiber orientation, both for accuracy and for computational speed.

\section{Fiber Motion Basics}
Jeffery's equation \cite{jeffery:23} has been used to predict the motion of the direction of axi-symmetric fibers under the influence of a low Reynold's number flow of a Newtonian fluid, whose velocity field is $\bu = \bu(\bx,t)$.  The directions of the fibers is represented by the fiber orientation distribution $\psi = \psi(\bx,\bp,t)$, where $\bp$ is an element of the orientation space, that is, the 2-dimensional sphere $S = \{\bp = (p_1,p_2,p_3):p_1^2+p_2^2+p_3^2=1\}$.  Thus given a subset $E$ of $S$, the proportion of fibers whose direction is in $E$ is given by $\int_E \psi(\bx,\bp,t) \, d\bp$, where $d\bp$ represents the usual integration over $S$.  In particular, an isotropic distribution is represented by $\psi = 1/4\pi$. The Jeffery's equation for the fiber orientation distribution is
\begin{equation}
\label{jeffery's}
\frac{D\psi}{Dt} = -\tfrac12 \pnabla\cdot((\Omega\cdot\bp + \lambda\Gamma\cdot\bp - \lambda \Gamma:\bp\bp\bp) \psi)
\end{equation}
Here $\Omega$ is the vorticity, that is, the anti-symmetric part $\bnabla\bu-(\bnabla\bu)^T$ of the Jacobian  of the velocity field $\bnabla\bu = (\partial u_i/\partial x_j)_{1\le i,j\le3}$, and $\Gamma$ is the rate of strain tensor, that is, the symmetric part $\bnabla\bu+(\bnabla\bu)^T$ of the Jacobean of the velocity field.  Also, $D/Dt = \partial/\partial t + \bu \cdot \bnabla$ represents the material derivative, and $\bnabla_\bp = (I-\bp\bp)\cdot\left(\tfrac\partial{\partial p_1},\tfrac\partial{\partial p_2},\tfrac\partial{\partial p_3}\right)$ is the gradient operator restricted to the sphere.

Equation~\eqref{jeffery's} is modified to incorporate the rotary diffusion expressed by Bird {\it et al.} \cite{bird:87b}, occasionally referred to as the generalized Fokker-Planck or the Smoluchowski equation \cite{petrie:99}, as
\begin{equation}
\label{jeffery's-folgar-tucker}
\frac{D\psi}{Dt} = -\tfrac12 \pnabla\cdot((\Omega\cdot\bp + \lambda\Gamma\cdot\bp - \lambda \Gamma:\bp\bp\bp) \psi) + \Delta_\bp(D_r \psi) ,
\end{equation}
where $D_r$ captures the effect of fiber interaction and depends upon the flow kinetics.  Here $\Delta_\bp = \bnabla_\bp\cdot\bnabla_\bp$ represents the Beltrami-Laplace operator on the sphere.  Folgar and Tucker \cite{folgar:84} selected $D_r = C_I \dot{\gamma}$ where $\dot{\gamma} = \left(\frac12\Gamma:\Gamma\right)^{1/2}$ and $C_I$ is a constant that depends upon the volume fraction and aspect ratio of the fibers.

Other authors have considered a wider class of diffusion terms.  For example, Koch \cite{koch:95}, and Phelps and Tucker \cite{phelps:09} considered anisotropic diffusion
\begin{equation}
\label{koch}
\frac {D\psi}{Dt} = -\tfrac12 \pnabla\cdot((\Omega\cdot\bp + \lambda\Gamma\cdot\bp - \lambda \Gamma:\bp\bp\bp) \psi) + \bnabla_\bp\cdot(I-\bp\bp)\cdot D_r\cdot\bnabla_\bp\psi
\end{equation}
where $D_r$ is the anisotropic diffusion matrix, calculated as a function of $\psi$ and $\bnabla u$ (see, e.g., \cite{phelps:09,koch:95}).

Since these are, in effect, partial differential equations in 5-spacial dimensions (3 for space and 2 for the orientation defined on a unit sphere), numerically calculating solutions can be rather daunting with solutions taking days to weeks for simple flows.  Hence Hinch and Leal \cite{hinch:73} suggested to recast the equation in terms of moment tensors.  For example, the second and fourth moment tensors are defined by
\begin{gather}
\label{2nd def}
A = \int_S \bp\bp\psi\,d\bp, \qquad \qquad \mathbb A = \int_S \bp\bp\bp\bp\psi\,d\bp
\end{gather}
Then Jeffery's equation~\eqref{jeffery's} for the second order moment tensor can be expressed as
\begin{equation}
\label{A-jeffery's}
\frac {DA}{Dt} = \tfrac12(\Omega\cdot A - A\cdot\Omega + \lambda(\Gamma\cdot A+A\cdot \Gamma) - 2 \lambda \mathbb A:\Gamma)
\end{equation}
and the equations~\eqref{jeffery's-folgar-tucker} and~\eqref{koch} with diffusion terms become
\begin{equation}
\label{A-jeffery's-diffusion}
\frac {DA}{Dt} = \tfrac12(\Omega\cdot A - A\cdot\Omega + \lambda(\Gamma\cdot A+A\cdot \Gamma) - 2 \lambda \mathbb A:\Gamma) + \mathcal D[A]
\end{equation}
where $\mathcal D[A]$ for isotropic diffusion as expressed in equation~\eqref{jeffery's-folgar-tucker} becomes
\begin{equation}
\label{D[A]-jeffery's-folgar-tucker}
\mathcal D[A] = D_r(2I-6A)
\end{equation}
and subsequently the anisotropic diffusion of equation~\eqref{koch} (see \cite{phelps:09}) is
\begin{equation}
\label{D[A]-jeffery's-koch}
\mathcal D[A] = 2D_r -2(\tr D_r)A - 5(A\cdot D_r+D_r\cdot A) + 10 \mathbb A:D_r
\end{equation}
The difficulty with equations~\eqref{A-jeffery's} and~\eqref{A-jeffery's-diffusion} is that they explicitly include the fourth order moment tensor, and implicitly the higher order diffusion models of equation~\eqref{D[A]-jeffery's-koch} include moments higher than the second-moment.  To circumvent this problem, various authors (for example, \cite{hinch:73,altan:89,altan:90,verleye:93,cintra:95,verweyst:99,chung:02,han:02,jack:05a}) have proposed \emph{closures}, that is, formulae to calculate the fourth order moment tensor $\mathbb A$ from the second order moment tensor $A$.  The mapping from $A$ to ${\mathbb A}$ is not unique, thus closures are only able to approximately obtain a higher order moment from the lower order moments. Most closures are often constructed by obtaining the best-fit coefficients of for a polynomial by fitting numerical data obtained by directly evaluating equation~\eqref{jeffery's-folgar-tucker} using a finite element method to solve equation~\eqref{jeffery's-folgar-tucker} (for example, Bay \cite{bay:91c}).

\section{The Fast Exact Closure}
Verleye and Dupret \cite{verleye:93} (see also \cite{wetzel:99b,verweyst:98,dinh:84,lipscomb:88,altan:93}) noted that there is an exact closure for Jeffery's equation when the diffusion terms are not present, \emph{in the particular case that the fiber orientation distribution is at some time isotropic}.  This exact closure is stated explicitly in \cite{montgomery-smith:10b} for the scenario when the suspension is dilute.  For the sake of labeling, the present closure retains the reference \emph{Exact Closure}, as it is exact for Jeffery's equation without diffusion terms.

The Exact Closure may be directly computed by solving the elliptic integral forms presented in equation~\eqref{A-jeffery's-diffusion}, where $\mathbb A$ is computed from $A$ using equations~\eqref{A(B)} and~\eqref{bbA} as derived in \cite{montgomery-smith:10b}. This approach only gives the exact answer to equations~\eqref{jeffery's-folgar-tucker} and~\eqref{koch} when $D_r=0$ and when the orientation is isotropic at some time.  Nevertheless it is reasonable to suppose that the exact closure should give a reasonable approximation in general, even when $D_r \ne 0$ as in Verweyst {\it et al.} \cite{verweyst:99,verweyst:98}.  Their ORT closure is a polynomial approximation to the Exact Closure, and as we demonstrate below, gives answers that are virtually indistinguishable from that of the Exact Closure.

The \emph{Fast Exact Closure} (FEC) performs the Exact Closure in a computationally efficient manner.  A version of FEC is described in \cite{montgomery-smith:10b}, but only when the diffusion terms are absent.  In this section we describe the FEC from an implementation perspective, and leave the full derivation to the appendix.

The idea behind the FEC is the computation of two rank 4 tensors $\mathbb C$ and $\mathbb D$, defined in equations~\eqref{C} and~\eqref{CDM}, respectively, which we define as \emph{conversion tensors}.  These tensors convert between $DA/Dt$ and $DB/Dt$ according to the formulae
\begin{equation}
\label{dadb-C}
\frac{DA}{Dt} = - \mathbb C:\frac{DB}{Dt}, \qquad \qquad
\frac{DB}{Dt} = - \mathbb D:\frac{DA}{Dt}
\end{equation}
as derived in equations~\eqref{proof_of_dadb-C} -~\eqref{S_of_dyad}. The orientation tensor $A$ retains the classical meaning as described in \cite{advani:87a} and the tensor $B$ turns out to be extremely useful for computations. $B$ appears to be a more abstract quantity to describe the degree of orientation much like the orientation tensor. For example, when the orientation parameter $B$ is given as $B_{ij} = \delta_{ij}$ this is analogous to saying that the orientation is isotropic, whereas when one of the diagonal terms of $B$ goes to $0$, it indicates that the orientation is perfectly aligned along the corresponding coordinate axis. Montgomery-Smith {\it et al.} \cite{montgomery-smith:10b} provide a further discussion as to the meaning of the orientation parameter $B$

What makes everything work is the formula, proven in the appendix by equation~\eqref{proof_of_CMB_symmetrix_matrix}, that for any matrix $M$, we have
\begin{equation}
\label{CBM}
\mathbb C:(B\cdot M+M^T\cdot B) = (\tr M) A + M\cdot A + A\cdot M^T - 2\mathbb A:M
\end{equation}
where $\mathbb A$ and $A$ satisfy equations~\eqref{A(B)} and~\eqref{bbA}.

The FEC present in this paper will be of the form:
\begin{equation}
\label{A-generic}
\frac {DA}{Dt} = - \mathbb C:F(B) + G(A), \qquad \qquad
\frac{DB}{D t} = F(B) - \mathbb D:G(A)
\end{equation}
where $F(B)$ and $G(A)$ will be given explicitly below. This is a general form that can be applied to a the known diffusion models that fit the form of equation~\eqref{jeffery's-folgar-tucker} or~\eqref{koch}. The conversion tensors $\mathbb C$ and $\mathbb D$ are defined later in this section, and in the appendix we provide a more mathematical formula for them along with a proof of the above properties. It is important to note that $\mathbb C$ and $\mathbb D$ may be computed directly from $A$ and $B$ in a rather fast manner, involving nothing more than the diagonalization and inversion of three by three symmetric matrices, general simple arithmetic, and where appropriate invoking inverse trigonometric or inverse hyperbolic functions.

The FEC solves the coupled ODEs of~\eqref{A-generic} simultaneously.  If the initial fiber orientation is isotropic, then  $A=\tfrac13 I$ and $B=I$ at $t=0$.  When the initial fiber orientation is not isotropic, then one can compute the initial condition for $B$ from $A$ by inverting equation~\eqref{A(B)}, as described in \cite{montgomery-smith:10b}.

It can be shown that the matrices $A$ and $B$ remain positive definite, simultaneously diagonalizable, and satisfy the equations $\tr A = \det B = 1$ for all time.

For example, the FEC for the Jeffery's equation with isotropic diffusion given in equation~\eqref{jeffery's-folgar-tucker} is given by:
\begin{gather}
\label{A}
\frac {DA}{Dt} = \tfrac12 \mathbb C:[B \cdot(\Omega+\lambda \Gamma) + (-\Omega+\lambda\Gamma)\cdot B] + D_r(2I-6A) \\
\label{B}
\frac{DB}{D t} =
- \tfrac12(B \cdot(\Omega+\lambda \Gamma) + (-\Omega+\lambda\Gamma)\cdot B) - D_r \mathbb D:(2I-6A)
\end{gather}
and the FEC\ for Jeffery's equation with anisotropic diffusion as shown in equation~\eqref{koch} is given by
\begin{gather}
\label{A-B-Koch}
\frac {DA}{Dt} = \tfrac12 \mathbb C:[B \cdot(\Omega+\lambda \Gamma) + (-\Omega+\lambda\Gamma)\cdot B]
+ 2D_r +3(\tr D_r)A - 5 \mathbb C:(B\cdot D_r + D_r\cdot B) \\
\label{B-A-Koch}
\frac{DB}{D t} =
- \tfrac12(B \cdot(\Omega+\lambda \Gamma) + (-\Omega+\lambda\Gamma)\cdot B)
 - \mathbb D:(2D_r +3(\tr D_r)A) + 5 (B\cdot D_r + D_r\cdot B)
\end{gather}
Using equation~\eqref{CBM} it can be seen that equation~\eqref{A} comes directly from equations~\eqref{A-jeffery's-diffusion} and~\eqref{D[A]-jeffery's-folgar-tucker}, and equation~\eqref{B} comes from applying equation~\eqref{CDM} to equation~\eqref{A}.  Similarly for the anisotropic diffusion model, this can be observed for equations~\eqref{A-B-Koch} and~\eqref{B-A-Koch}.

Notice, for equations~\eqref{A} and~\eqref{B} and for equations~\eqref{A-B-Koch} and~\eqref{B-A-Koch}, that the fourth-order orientation tensor ${\mathbb A}$ does not appear. The equation of motion for the orientation is now reduced to developing the relationship between $A$ and $B$ with that of ${\mathbb C}$ and ${\mathbb D}$. The conversion tensors ${\mathbb C}$ and ${\mathbb D}$ are both computed with respect to the basis of orthonormal eigenvectors of $B$.  With respect to this basis, the matrix $B$ is diagonal with entries $b_1$, $b_2$ and $b_3$, and $A$ is diagonal with entries $a_1$, $a_2$ and $a_3$ where we constrain $b_1 \le b_2 \le b_3$ which implies that $a_1 \ge a_2 \ge a_3$.

If the eigenvalues $b_1$, $b_2$ and $b_3$ are not close to each other, then $\mathbb C$ is the symmetric tensor calculated using the formulae from equations~\eqref{C long} and~\eqref{C:I} from the appendix
\begin{equation}
\label{C quick}
\begin{array}{lll}
\mathbb C_{1122} = \frac{a_{1} - a_{2}}{2(b_2-b_1)} & \qquad & \mathbb C_{1111} = \tfrac12 b_1^{-1} - \mathbb C_{1122} - \mathbb C_{1133} \\
\mathbb C_{1133} = \frac{a_{1} - a_{3}}{2(b_3-b_1)} & \qquad & \mathbb C_{2222} = \tfrac12 b_2^{-1} - \mathbb C_{1122} - \mathbb C_{2233} \\
\mathbb C_{2233} = \frac{a_{2} - a_{3}}{2(b_3-b_2)} & \qquad & \mathbb C_{3333} = \tfrac12 b_3^{-1} - \mathbb C_{1133} - \mathbb C_{2233} \\
\mathbb C_{ijkk} = 0 \text{ if $i\ne j\ne k$}
\end{array}
\end{equation}
If two or more of the eigenvalues are close to each other, then these equations can give rise to large numerical errors, or even `divide by zero' exceptions.  So in this situation, we use different formulae to compute $\mathbb C$.

Suppose two of the eigenvalues are close to each other, for example, $b_1 = b_0+\epsilon$ and $b_2 = b_0-\epsilon$, where $\epsilon$ is small.  Thus $b_0 = \tfrac12(b_1+b_2)$ and $\epsilon = \tfrac12(b_1-b_2)$.  Define the quantity $\mathcal I_n$ from equation~\eqref{In def} and with equations~\eqref{In expansion} and~\eqref{In expansion2} this quantity can be expressed as
\begin{equation}
\label{In}
\begin{split}
\mathcal I_{n+1} = \frac{2 n -1}{2 n (b_0-b_3)} \mathcal I_n - \frac{\sqrt{b_3}}{n b_0^n (b_0-b_3)} \text{ if $n\ge 1$} \\
\mathcal I_1 = \frac{2}{\sqrt{b_0-b_3}}\cos^{-1}\left(\sqrt{\frac{b_3}{b_0}}\right) \text{ if $b_0>b_3$} \\
\mathcal I_1 = \frac{2}{\sqrt{b_3-b_0}} \cosh^{-1}\left(\sqrt{\frac{b_3}{b_0}}\right) \text{ if $b_0<b_3$}
\end{split}
\end{equation}
Then replace the first equation of equation~\eqref{C quick} by
\begin{equation}
\label{C quick b1=b2+epsilon}
\mathbb C_{1122} = \tfrac14 \mathcal I_3 + \tfrac38 \mathcal I_5 \epsilon^2 + O(\epsilon^4)
\end{equation}
If all three of the eigenvalues are almost equal, that is $b_1=1+c_1$, $b_2=1+c_2$, $b_3=1+c_3$ with $|c_1|,|c_2|,|c_3|\le\epsilon$, then it can be similarly shown that
\begin{equation}
\label{C quick b1=b2=b3+epsilon}
\begin{split}
\mathbb C_{1122} &= \textstyle \frac{1}{10}-\frac{3}{28} c_1-\frac{3}{28}c_2-\frac{1}{28}c_3
   +\frac{5}{48}c_1^2+\frac{1}{8}c_1c_2+\frac{1}{24}c_1c_3+\frac{5}{48} c_2^2+\frac{1}{24}c_2c_3+\frac{1}{48}c_3^2 \\
& \phantom{={}}
\textstyle -\frac{35}{352}c_1^3-\frac{45}{352}c_1^2c_2-\frac{15}{352}c_1^2c_3-\frac{45}{352}c_1c_2^2-\frac{9}{176}c_1c_2c_3 \\
& \phantom{={}}
\textstyle
   -\frac{9}{352}c_1c_3^2-\frac{35}{352}c_2^3-\frac{15}{352}c_2^2c_3-\frac{9}{352}c_2c_3^2-\frac{5}{352}c_3^3 + O(\epsilon^4)
\end{split}
\end{equation}
with similar formulae for $\mathbb C_{1133}$ and $\mathbb C_{2233}$.  The remaining entries of $\mathbb C$ are computed using the last four equations from~\eqref{C quick}.

The rank 4 conversion tensor $\mathbb D$ given in equation~\eqref{dadb-C} is defined through equation~\eqref{CDM} with respect to the basis of orthonormal eigenvectors of $B$, and can be simplified to
\begin{equation}
\label{D}
\begin{split}
\left[\begin{smallmatrix}
\mathbb D_{1111} & \mathbb D_{1122} & \mathbb D_{1133} \\
\mathbb D_{2211} & \mathbb D_{2222} & \mathbb D_{2233} \\
\mathbb D_{3311} & \mathbb D_{3322} & \mathbb D_{3333}
\end{smallmatrix}\right]
\quad &= \quad
\left[\begin{smallmatrix}
\mathbb C_{1111} & \mathbb C_{1122} & \mathbb C_{1133} \\
\mathbb C_{2211} & \mathbb C_{2222} & \mathbb C_{2233} \\
\mathbb C_{3311} & \mathbb C_{3322} & \mathbb C_{3333}
\end{smallmatrix}\right]^{-1} \\
\mathbb D_{ijij} = \mathbb D_{ijji} = \frac1{4\mathbb C_{ijij}} \text{ if $i\ne j$} & \qquad \quad
\mathbb D_{ijkk} = 0 \text{ if $i\ne j\ne k$}
\end{split}
\end{equation}
Note that there is no reason to suppose that $\mathbb D$ is completely symmetric because in general $\mathbb D_{ijij}$ will not be the same as $\mathbb D_{iijj}$.

In performing the numerical calculations, it is more efficient when forming $DA/Dt$ and $DB/Dt$ from equation~\eqref{A-generic} to calculate the right hand side in the coordinate system of the orthonormal eigenvectors of $B$, and then convert back to the standard coordinate system when solving for $A$ and $B$.

For example, suppose $\mathbb B$ is any rank four tensor such that $\mathbb B_{ijkk}=0$ if $i\ne j\ne k$, and $\mathbb B_{ijkl} = \mathbb B_{jikl} = \mathbb B_{klij}$.  Suppose also that $N$ is a symmetric matrix.  Then $\mathbb B:N$ can be calculated by first defining the matrices $M_{\mathbb B}$ and $\tilde M_{\mathbb B}$ as
\begin{equation}
M_{\mathbb B} = \left[\begin{smallmatrix}\mathbb B_{1111} & \mathbb B_{1122} & \mathbb B_{1133} \\ \mathbb B_{1122} & \mathbb B_{2222} & \mathbb B_{2233} \\ \mathbb B_{1133} & \mathbb B_{2233} & \mathbb B_{3333}\end{smallmatrix}\right] ,
\qquad
\tilde M_{\mathbb B} = \left[\begin{smallmatrix}0 & \mathbb B_{1212} & \mathbb B_{1313} \\ \mathbb B_{1212} & 0 & \mathbb B_{2323} \\ \mathbb B_{1313} & \mathbb B_{2323} & 0\end{smallmatrix}\right]
\end{equation}
then decompose
\begin{equation}
N = \text{diag}(\mathbf n) + \tilde N
\end{equation}
where $\mathbf n = (N_{11},N_{22},N_{33})$, and $\tilde N$ is the matrix of the off-diagonal elements of $N$.  It follows that
\begin{equation}
\label{B on N}
\mathbb B:N = \text{diag}(M_{\mathbb B}\cdot \mathbf n) + 2 \tilde M_\mathbb B \circ \tilde N
\end{equation}
where for any matrices $U$ and $V$ we define the entrywise product (also known as the Hadamard or Schur product) by $(U\circ V)_{ij} = U_{ij} V_{ij}$.

\subsection{The Reduced Strain Closure}
Wang {\it et al.} \cite{wang:08} described a method that slows down the rate of alignment of the fibers, which the paper calls the reduced strain closure model (RSC).  The method is implemented by selecting a number $0<\kappa\le1$, which is identified as the rate of reduction.  The authors \cite{wang:08} define the tensor
\begin{equation}
\mathbb M = \sum_{i=1}^3 \e_i\e_i\e_i\e_i
\end{equation}
where $\e_1$, $\e_2$, $\e_3$ are the orthonormal eigenvectors for $A$.  The RSC replaces equations of the form
\begin{equation}
\label{DADt-generic}
\frac{DA}{Dt} = F(A)
\end{equation}
by
\begin{equation}
\label{DADt-generic-rsc}
\frac{DA}{Dt} = F(A) - (1-\kappa)\mathbb M:F(A)
\end{equation}
It turns out this form is simple to reproduce for the FEC.  If equation~\eqref{DADt-generic} is represented by the FEC
\begin{equation}
\label{FEC-A-generic}
\frac{DA}{Dt} = F(A,B), \qquad \frac{DB}{Dt} = G(A,B)
\end{equation}
then the effect of equation~\eqref{DADt-generic-rsc} is precisely modeled by the new FEC
\begin{equation}
\label{FEC-A-generic-rsc}
\frac{DA}{Dt} = F(A,B) - (1-\kappa)\mathbb M:F(A,B), \qquad
\frac{DB}{Dt} = G(A,B) - (1-\kappa)\mathbb M:G(A,B)
\end{equation}
Finally, from a computational point of view, it should be noticed that if we are working in the basis of orthonormal eigenvectors of $B$, then for any symmetric matrix $N$ we have that $\mathbb M:N$ is simply the diagonal part of $N$, that is, $\text{diag}(N_{11},N_{22},N_{33})$.

\subsection{Is the solution to FEC always physical?}
By the phrase ``the solutions stay physical'' we mean that $A$ stays positive definite with trace one, that is, there exists a fiber orientation distribution $\psi$ that satisfies equation~\eqref{2nd def}.  In fact, if $A$ ever ceases to become positive definite, then not only is the Exact Closure going to give the wrong answer, it even ceases to have a meaning in that equation~\eqref{A(B)} which is used to define $A$ in terms of $B$ cannot be solved.  Thus another way to state ``the solutions stay physical'' is that $B$ stays positive definite and finite, that is, none of the eigenvalues of $B$ become zero, and none of them become infinite.

\begin{thm}
\label{thm-ft}
The FEC solution to the isotropic diffusion equations~\eqref{A} and~\eqref{B} have global in time physical solutions if $\Omega$, $\Gamma$ and $D_r$ are bounded.
\end{thm}

\begin{thm}
\label{thm-Koch}
The FEC solution to the anisotropic diffusion equations~\eqref{A-B-Koch} and~\eqref{B-A-Koch} have global in time physical solutions if $D_r$ is positive definite, and $\Omega$, $\Gamma$, $D(D_r)/Dt$, $D_r$ and $1/\|D_r^{-1}\|$ are bounded.
\end{thm}

\noindent
where the proofs for both theorems are given in the Appendix beginning with equation~\eqref{DAB}. Unfortunately Theorem~\ref{thm-Koch} will not necessarily apply to the Koch model \cite{koch:95} nor to the Phelps-Tucker ARD model \cite{phelps:09}, as there is no guarantee that $1/\|D_r^{-1}\|$ is bounded nor, in the ARD case, that $D_r$ is positive definite, unless extra hypotheses are applied.

\subsection{Algorithm Summary}
\label{subsec:algorithm_summary}
The algorithm to solve the FEC closure for the second-order orientation tensor $A$ and the second-order tensor $B$ can be summarized as:
\begin{enumerate}
\item
Initialize $A$ and $B$, and define $\lambda$ along with any constants needed for the diffusion model ${\mathcal D}\left[A\right]$
\item
At time $t_i$, rotate the tensors $A$ and $B$ into the principal frame of $B$
\item
When the eigenvalues are distinct, use equation~\eqref{C quick} for ${\mathbb C}$.  Otherwise when two eigenvalues are repeated, use equation~\eqref{In} along with equation~\eqref{C quick b1=b2+epsilon}, or in the case when three eigenvalues are repeated, use equation~\eqref{C quick b1=b2=b3+epsilon}.
\item
From ${\mathbb C}$, compute ${\mathbb D}$ using equation~\eqref{D} in the principal frame of $B$
\item
Compute $DA/Dt$ and $DB/Dt$ using either equations~\eqref{A} and~\eqref{B} for isotropic diffusion or equations~\eqref{A-B-Koch}, \eqref{B-A-Koch} and~\eqref{FEC-A-generic-rsc} for the anisotropic diffusion model, ARD-RSC.  For the symmetric rank four tensor contractions with rank two tensors, use equation~\eqref{B on N} to reduce the number of redundant multiplication operations.
\item
Rotate $DA/Dt$ and $DB/Dt$ into the flow reference frame, and extrapolate $A\left(t_{i+1}\right)$ and $B\left(t_{i+1}\right)$ from time $t_i$ using any standard ODE solver.
\end{enumerate}

There are a number of coding issues we encountered, and we feel it will be helpful to share as it will aid others in their computational implementations.
\begin{itemize}
\item
There is a choice to compute the basis of orthonormal eigenvectors from either $A$ or $B$, where in theory these should be identical. We compute the basis from $B$, arguing that the quantity $B$ is somehow more `fundamental' and $A$ is `derived' from $B$,  which is true in the absence of diffusion.
\item
We solve a ten dimensional set of ODEs, five for $A$, and five for $B$, where one of the components of both $A$ and $B$ can be obtained, respectively, from the relationships $\tr A =1$ and $\det B = 1$.
\item
When computing $A$ from the orthonormal eigenvector basis of $B$, it is important to force the off diagonal entries to be non-zero to limit numerical drifting.  In our studies, we found that failing to do this could cause an adaptive ODE solver to completely freeze in select scenarios.
\item
We set the ODE solver to work with a relative tolerance of $10^{-5}$, and choose to use equations~\eqref{C quick b1=b2+epsilon} or~\eqref{C quick b1=b2=b3+epsilon} when the eigenvalues were within $10^{-4}$ of each other.  This should cause $\mathbb C$ to be computed with an accuracy of about $10^{-8}$ when using equations~\eqref{C quick}, and nearly machine precision when using equations~\eqref{C quick b1=b2+epsilon} or~\eqref{C quick b1=b2=b3+epsilon}.
\end{itemize}

\section{Numerical Results}
Results are presented to demonstrate the accuracy improvements from employing the FEC closure, and just as important to demonstrate the computational speed advances over the similarly accurate orthotropic closures. In the present examples, all flows have an initial isotropic orientation state designated by $A_{11}=A_{22}=A_{33}=1/3$ and $B_{11}=B_{22}=B_{33}=1$, with all other components of $A$ and $B$ being zero. The accuracy of the closure does not depend on the initial orientation state, the isotropic orientation state is chosen for uniformity. The equations of motion are solved using the FEC closure for $A$ and $B$ from equations~\eqref{A} and~\eqref{B} for isotropic diffusion or from equations~\eqref{A-B-Koch}, \eqref{B-A-Koch} and~\eqref{FEC-A-generic-rsc} for the anisotropic rotary diffusion model with the reduced strain closure ARD-RSC from Phelps and Tucker \cite{phelps:09}. For comparison, the classical equations of motion for the second-order orientation tensor $A$ requiring a curve-fitted closure for the fourth-order orientation tensor ${\mathbb A}$, are solved using equations~\eqref{A-jeffery's-diffusion} and~\eqref{D[A]-jeffery's-folgar-tucker} for Folgar-Tucker diffusion and equations~\eqref{A-jeffery's-diffusion}, \eqref{D[A]-jeffery's-koch} and~\eqref{DADt-generic} for the ARD-RSC diffusion model. Results are compared to solutions obtained using the Spherical Harmonic approach \cite{montgomery:10} for solving the full distribution function equations~\eqref{jeffery's-folgar-tucker} and~\eqref{koch}. It has been demonstrated in \cite{montgomery:10} that solutions using the Spherical Harmonic approach are only limited in their accuracy by machine precision and require considerably less computational effort than solutions using the control volume approach of Bay \cite{bay:91c}. Although a great reduction in speed and an advancement in accuracy, the Spherical Harmonic approach still requires more effort than the orientation tensor approach, nor does it readily lend itself to an applicable form for coupling with commercial FEA solvers. We select three commonly employed closures for comparisons. The first is the classical Hybrid closure of Advani and Tucker \cite{advani:87a} is selected as it is regularly used in commercial and research codes due to its computational efficiency and ease of implementation. The second is an orthotropic closure, whose class of closures has found increasing use due to their considerable accuracy improvements over the Hybrid closure.  In our study we select the ORT closure presented by VerWeyst and Tucker \cite{verweyst:99} based on the Wetzel closure \cite{wetzel:99b}.  Our third closure is that of the IBOF from Chung and Kwon \cite{chung:02} which is claimed to be a more computationally efficient orthotropic closure as it uses the invariants of $A$ as opposed to the eigenvalues of $A$ thus avoiding costly tensor rotations.

\subsection{Results: Simple Shear Flow}
\label{subsec:results_SSFlow}
The first example is that of a pure shearing flow, given by $v_1=Gx_3$ and $v_2=v_3=0$.  Pure shearing flow is commonly employed (see e.g., \cite{cintra:95,chung:02,jack:10a}) to demonstrate a particular closure problem due to the oscillatory nature of alignment inherent to the Jeffery fiber orbits. Two scenarios are presented, the first of the Folgar-Tucker isotropic diffusion model in equation~\eqref{jeffery's-folgar-tucker} where $D_r=C_I\dot{\gamma}$, and the second scenario for the ARD-RSC anisotropic diffusion model.

\subsubsection{Simple Shear Flow Orientation}
\label{subsubsec:results_SSFlow_Orientation}
In industrial simulations, the Folgar-Tucker isotropic diffusion model typically has interaction coefficients that range from $C_I=10^{-3}$ to $C_I=10^{-2}$. The effective fiber aspect ratio ranges from 5 to 30 ($a_e \simeq 1.4 \times a_r$, where $a_r$ is the aspect ratio of cylindrical fibers), which corresponds to a shape correction factor ranging from $\lambda = 0.96$ to $\lambda = 0.999$. Two simulation results using isotropic diffusion are presented in Figures \ref{fig:Simple_Shear_FT}(a) and (b), the first is for $C_I=10^{-3}$ with $\lambda = 0.99$ and the later for $C_I=10^{-2}$ with $\lambda = 0.95$. Results for the IBOF closure are not shown as they are nearly graphically indistinguishable from the ORT closure results. It is important to observe that the ORT and the FEC closure yield results that are graphically indistinguishable and reasonably close to the orientation state predicted from the numerically exact Spherical Harmonic solution.  Conversely, the orientation results from the Hybrid closure tend to over predict the the true orientation state. It is important to point out the apparent oscillatory nature of the transient solution for the Spherical Harmonic results when $C_I=10^{-3}$ with $\lambda=0.99$, which occurs to a lesser extent for $C_I=10^{-2}$. These oscillations are expected due to the low amount of diffusion present. Equally important is to notice that the oscillations from the FEC closure, as well as the ORT, both damp out to the same steady state value. Note also that the FEC does not oscillate excessively for either of the isotropic flow conditions presented, which was a problem that plagued the early orthotropic closures (see e.g., \cite{cintra:95} and \cite{chung:01}) and the early neural network closures \cite{qadir:09}. There remains room for further accuracy improvements (see e.g., \cite{mullens:10} for several preliminary higher accuracy closures). However, it is speculated based upon the discussion in Jack and Smith \cite{jack:04a} that such improvements will be slight when solving the second-order moment equations, and higher order moment simulations, such as those that use sixth-order closures (see e.g., \cite{jack:05a}) may need to be considered for significant accuracy improvements.

\begin{figure}[p]
\begin{center}
\includegraphics[height=2.5in]{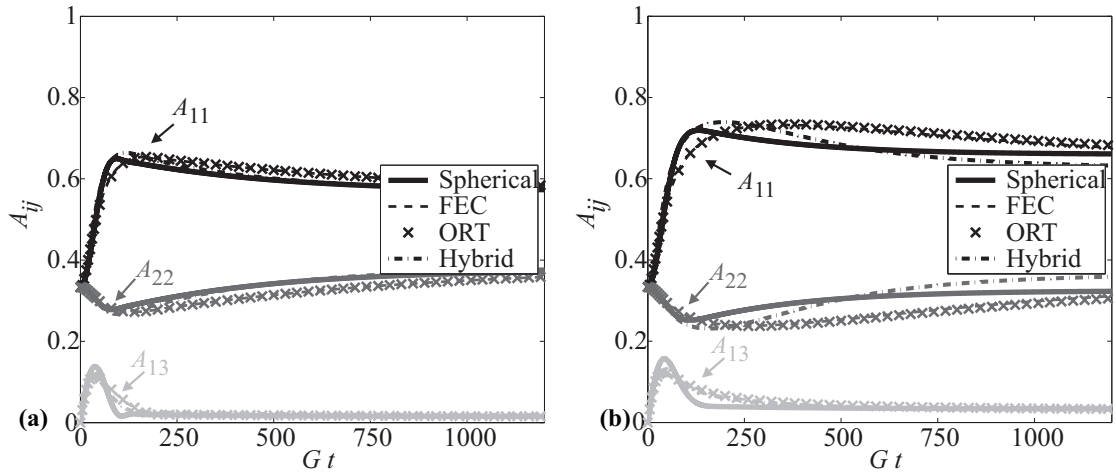}
\caption{Transient Solution for selected components of $A$ for simple shear flow under anisotropic rotary diffusion (ARD-RSC) (a) $\kappa=1/30$ and $\lambda = 0.95$ and (b) $\kappa=1/30$ and $\lambda = 1.0$.}
\label{fig:Simple_Shear_ARD}
\end{center}
\end{figure}

The Folgar-Tucker model has been used for decades, but tends to overstate the rate of alignment during the transient solution (see e.g., \cite{tucker:04}). The ARD-RSC model \cite{phelps:09} seeks to address these limitations, but few studies have focused on this new diffusion model and the dependance of computed results on the choice of closure.  In the ARD-RSC model, the rotary diffusion coefficient of Folgar and Tucker isotropic diffusion model ($D_r = C_I \dot{\gamma}$ where $\dot{\gamma} = \left(\frac12\Gamma:\Gamma\right)^{1/2}$) is replaced by an anisotropic diffusion coefficient  expressed by
\begin{equation}
D_r =  b_1 \dot{\gamma} I + b_2 \dot{\gamma} A + b_3  \dot{\gamma} A^2 + \tfrac12 b_4 \Gamma + \tfrac14 b_5\dot{\gamma}^{-1} \Gamma^2 \end{equation}
where
\begin{equation}
(b_1,b_2,b_3,b_4,b_5) = (1.924 \times 10^{-4},5.839\times 10^{-3} ,4.0\times 10^{-2},1.168\times10^{-5},0) \end{equation}
The ARD-RSC model serves as an excellent example of the effectiveness of the FEC approach for solving the tensor form of orientation as the ARD-RSC model will yield orientation states that are considerably different than that of the Folgar-Tucker model. Results from the various closures and the spherical harmonic results are presented in Figure~\ref{fig:Simple_Shear_ARD} for the ARD-RSC flow with $\kappa=1/30$. The value of $\kappa =1/30$ is taken from the results presented in Phelps and Tucker \cite{phelps:09}, which was based on their experimental observations. For a fiber aspect ratio of $\sim 5$, corresponding to $\lambda=0.95$, each of the investigated closures produces graphically similar results. During the initial flow stages, the Hybrid tends to over predict alignment, whereas the ORT and the FEC tend to under predict alignment.  As steady state is attained, the FEC and the ORT yield nearly identical results, both of which over predict $A_{11}$ in the final orientation state whereas the Hybrid yields a reasonable representation of the orientation. For a long fiber, corresponding to $\lambda\rightarrow 1$, the trends are similar to those of the lower aspect ratio fibers, but in this case the FEC and the ORT better represent the final orientation state relative to the Hybrid.

\subsubsection{Orthotropic Closure Errors}
\label{subsubsec:Orthotropic_Closure_Errors}
\begin{figure}[p]
\begin{center}
\includegraphics[height=2.5in]{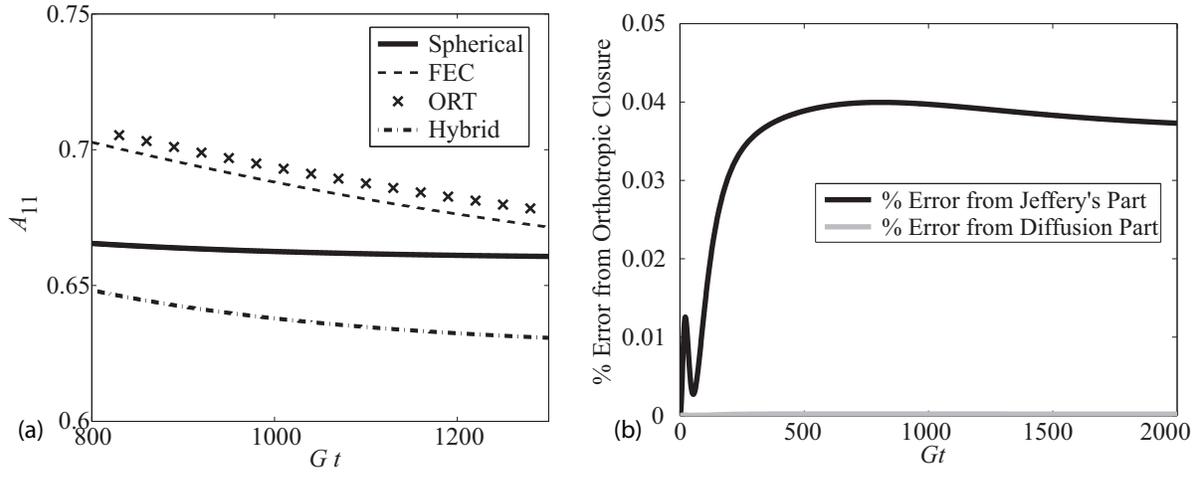}
\caption{Anisotropic rotary diffusion results, simple shear $\kappa=1/30$ and $\lambda = 1.0$ (a) Selected time range of for $A_{11}$ (b) Transient error in derivative computation for the fitted orthotropic closure ORT compared to FEC.}
\label{fig:Simple_Shear_ARD_lambda_1_00_Error}
\end{center}
\end{figure}

The ORT is a polynomial approximation to the Exact Closure, as demonstrated in the preceding section, and it is not surprising that the two approaches yield graphically indistinguishable results for many of the flows investigated. On closer inspection of the transient solution of the ARD-RSC model for $\kappa=1/30$ and $\lambda=1$ there is a slight difference.  This difference is shown in Figure~\ref{fig:Simple_Shear_ARD_lambda_1_00_Error}(a) where a closeup view is provided of the $A_{11}$ component for the flow times of $800\le Gt \le 1,200$. These results indicate how well the fitting was performed in the construction of the ORT. As the ORT is an approximation of the Exact Closure of Montgomery-Smith {\it et al.} \cite{montgomery-smith:10b} for pure Jeffery's flow, it is of interest to determine whether the slight deviation comes from the Jeffery's component or the diffusion component of equation~\eqref{A-jeffery's-diffusion}. To this end, we performed a comparison for the derivative of $A$ computed in two different ways. First, for each point in time $t$, we computed $A(t)$ and $B(t)$ using the FEC method.  Then we computed four quantities: $\frac{DA^{\text{\tiny FEC, Diff}}}{Dt}$ which contains the terms from the right hand side of equation~\eqref{A-B-Koch} that explicitly include $D_r$, $\frac{DA^{\text{\tiny FEC, Jeff}}}{Dt}$ which contains the terms from the right hand side of equation~\eqref{A-B-Koch} that do not involve $D_r$, $\frac{DA^{\text{\tiny ORT, Diff}}}{Dt}$ the right hand side of equation~\eqref{D[A]-jeffery's-koch}, and $\frac{DA^{\text{\tiny ORT, Jeff}}}{Dt}$ the right hand side of equation~\eqref{A-jeffery's-diffusion} when $\mathcal D(A)$ is set to zero. In the latter two cases $\mathbb A$ is computed using the ORT closure. The error is then defined as
\begin{eqnarray}
\label{error_ORT_with_FEC_Diffusion}
E_{\text{\tiny Diffusion}} = \sqrt{
\sum_{i=1}^3 \sum_{j=1}^3 \left(  \frac{DA_{ij}^{\text{\tiny FEC, Diff}}}{Dt}  -  \frac{DA_{ij}^{\text{\tiny ORT, Diff}}}{Dt}  \right)^2 }
\end{eqnarray}
\begin{eqnarray}
\label{error_ORT_with_FEC_Jeffery}
E_{\text{\tiny Jeffery}} = \sqrt{
\sum_{i=1}^3 \sum_{j=1}^3 \left(  \frac{DA_{ij}^{\text{\tiny FEC, Jeff}}}{Dt}  -  \frac{DA_{ij}^{\text{\tiny ORT, Jeff}}}{Dt}  \right)^2 }
\end{eqnarray}
Each of the two errors are plotted in Figure~\ref{fig:Simple_Shear_ARD_lambda_1_00_Error}(b). It is clear from the figure that although the ORT's derivative calculation from the diffusion component is not zero, it is minor in comparison to the error from the Jeffery's part of the orientation tensor equation of motion.  This error is only a rough indication of the sources of error, but values of 0.04\% at a given moment in flow time can account for an error as large as 40\% for $A$ for the flow times on the order 1,000. Since the errors from each of the possible sources probably do not drive the error in the solution toward the same direction, the total error would be expected to be less than the upper bound of 40\%, where in reality the error is closer to 0.9\% as steady state is approached.

Since the ORT and FEC differ by about 0.9\%, it begs the question as to which is more accurate in computing the true exact closure.  While the FEC in theory should exactly compute the exact closure, it is possible that numerical errors creep into the FEC.  To test for this, we performed a consistency check.  After finding the solution $A(t)$ and $B(t)$ using the FEC, we calculated
\begin{equation}
E_{\text{\tiny Exact}} = \sqrt{\sum_{i=1}^2 \sum_{j=1}^3 \left(A(B)_{ij}-A_{ij}\right)^2}
\end{equation}
where $A(B)$ was computed using equation~\eqref{A(B)}.  This calculation was performed by diagonalizing $B$, applying the elliptic integrals in equation set~\eqref{A long} using the software package \cite{gsl}, and then performing the reverse change of basis. The results for the ARD-RSC model with $\kappa = 1/30$ and $\lambda = 1.00$ show an error of less than $10^{-8}$ throughout the transient solution, thus suggesting the implementation as presented in this paper for the FEC is quite accurate.

\subsection{Results: Orientation Error Summary}
\label{subsec:results_Error}
\begin{figure}[p]
\begin{center}
\includegraphics[height=2.5in]{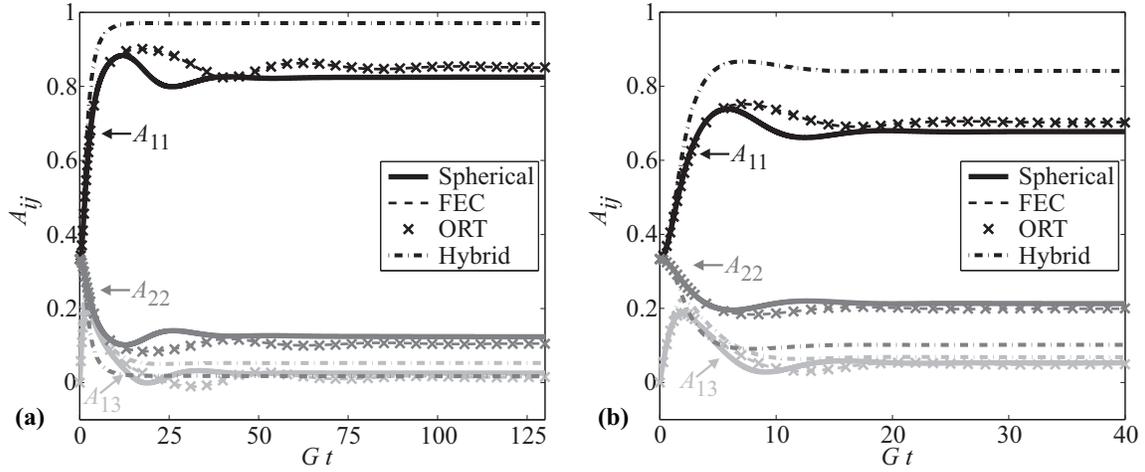}
\caption{Transient Solution for selected components of $A$ for simple shear flow under isotropic diffusion (a) $C_I=10^{-3}$ and $\lambda = 0.99$ and (b) $C_I=10^{-2}$ and $\lambda = 0.95$.}
\label{fig:Simple_Shear_FT}
\end{center}
\end{figure}

To quantify the errors observed in Figures \ref{fig:Simple_Shear_FT}(a) and (b) for the isotropic diffusion models, a series of fourteen flows are studied as outlined in table~\ref{norm_error_table} where $\lambda = 1$ for each of the flows. The solution is obtained using the classical closure methods and the FEC closure results are compared to solutions obtained from the Spherical Harmonic approach. To quantify the error, the time average of the Frobenius Norm of the difference between the true solution $A_{ij}^{\mbox{\tiny Spherical}}(t)$ and the approximate solution obtained from a closure $A_{ij}^{\mbox{\tiny Closure}}(t)$ is computed as
\begin{eqnarray}
\label{Avg_Frobenius_Norm}
\overline{E}_{\mbox{\tiny Closure}} = \frac{1}{t_{f}-t_{0}} \int_{t_{0}}^{t_{f}}\sqrt{\sum_{i=1}^3\sum_{j=1}^3\left|A_{ij}^{\mbox{\tiny Spherical}}(t) - A_{ij}^{\mbox{\tiny Closure}}(t)\right|^2}dt
\end{eqnarray}
where $t_{0}$ is the initial time where the fiber orientation is isotropic and $t_{f}$ is the time when the steady state is attained, which in this example will be defined when the magnitude of the largest derivative of the eigenvalues of $A$ is less than $G\times10^{-4}$.  This can be expressed as the smallest moment in time when the following is satisfied $\left(\max_{i\in\{1,2,3\}} |\frac{DA_{(i)}}{Dt}(t)|\right)\leq G\times10^{-4}$.  The quantitative error metric in equation~\eqref{Avg_Frobenius_Norm} yields a value for the simple shear flow of Figure \ref{fig:Simple_Shear_FT}(b) for the FEC, ORT and Hybrid closures of, respectively, $4.74\times10^{-2}$, $4.85\times10^{-2}$ and $1.75\times10^{-1}$. As the objective is to compare the relative accuracy improvements between the FEC closure and the existing closures we will normalize the error metric in equation~\eqref{Avg_Frobenius_Norm} as
\begin{eqnarray}
\label{norm_error}
\overline{\varepsilon}_{\mbox{\tiny Closure}} & \equiv & \frac{\overline{E}_{\mbox{\tiny Closure}}} {\min\limits_{\mbox{\tiny Closure}}\left(\overline{E}_{\mbox{\tiny Closure}}\right)}
\end{eqnarray}
where the closure with the greatest accuracy will have a value of $\overline{\varepsilon}_{\mbox{\tiny Closure}}=1$, and the remaining closures will have a value of $\overline{\varepsilon}_{\mbox{\tiny Closure}}$ in excess of 1.  For each of the flows studied, the normalized error of equation~\eqref{norm_error} is tabulated in Table \ref{norm_error_table} for the FEC, ORT, IBOF and the Hybrid closures. In each of the flows considered, the FEC performs as well as or better than the orthotropic closures.

\subsection{Results: Combined Flow}
\label{subsec:results_Combined_Flow}
\begin{figure}[p]
\begin{center}
\includegraphics[height=2.5in]{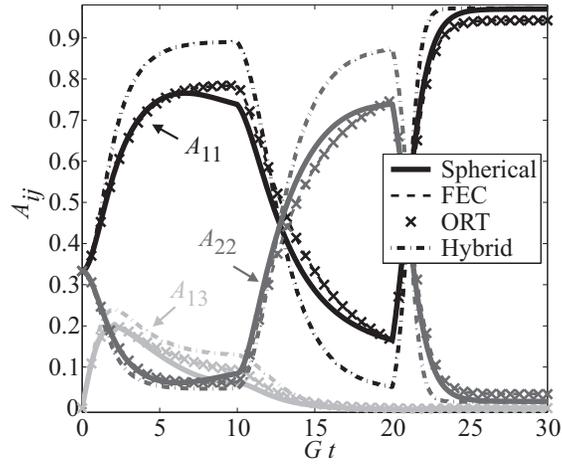}
\caption{Transient solution for selected components of $A$ for mixed flow from the Folgar-Tucker model with $C_I=10^{-2}$ and $\lambda = 1.0$.}
\label{fig:Mixed_FT_CI_0_01_lambda_1_00}
\end{center}
\end{figure}

A classical flow to demonstrate the effectiveness and robustness of a closure is that of the combined flow presented in Cintra and Tucker \cite{cintra:95}.  This flow is often selected as the orientation state crisscrosses the eigenspace of possible orientations. The combined flow begins with pure shear in the $x_1-x_2$ direction for $0\leq Gt < 10$ defined by the velocity field $v_1=Gx_2$, $v_2=v_3=0$.  The flow then transitions to shearing flow in the $x_2-x_3$ plane with stretching in the $x_3$ direction during the time $10\leq Gt<20$ defined by the velocity field $v_1=-1/20Gx_1$, $v_2=-1/20Gx_2+Gx_3$ and $v_3=1/10Gx_3$.  The flow then transitions to a flow with a considerable amount of stretching in the $x_1$ direction with a reduced amount of shearing in the $x_2-x_3$ plane for $20\leq Gt$ defined by the velocity field $v_1=Gx_1$, $v_2=-1/2Gx_2+Gx_1$ and $v_3=-1/2x_3$. The times where the flow transitions are chosen to prevent the orientation from attaining steady state, thus any error in the transient solution will be propagated to the next flow state.  As observed in Figure~\ref{fig:Mixed_FT_CI_0_01_lambda_1_00} for flow results from the Folgar-Tucker model with $C_I=10^{-2}$ and $\lambda=1$, the ORT and the FEC again yield similar results. This is significant as it further demonstrates the robustness and the accuracy of the FEC.

\subsection{Results: Center-gated Disk Flow}
\label{subsec:results_Disk_Flow}
\begin{figure}[p]
\begin{center}
\includegraphics[height=2.5in]{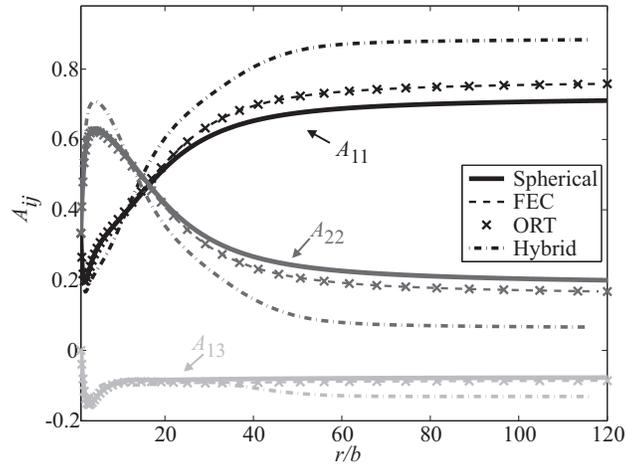}
\caption{Transient solution for selected components of $A$ for center-gated disk flow from the Folgar-Tucker model with $C_I=10^{-2}$ and $\lambda = 1.0$ for $r/b=4/10$.}
\label{fig:disc_zb_0_4_CI_0_01_FEC}
\end{center}
\end{figure}

The final flow investigated is that of the center-gated disk, a typical flow condition in industrial processes \cite{verweyst:99,jack:08a}. The flow enters the mold through the pin gate and flows radially outward, where the velocity is a function of both the gap height $2b$ and the radial distance from the gate $r$. The velocity gradient for a Newtonian fluid can be represented by \cite{cintra:95}
\begin{eqnarray}
\label{disk_vel}
v_r=\frac{3Q}{8\pi rb}\left(1-\left(\frac{z}{b}\right)^2\right),\:\:\:\: v_\theta=v_z=0
\end{eqnarray}
\begin{eqnarray}
\label{radialvelocitygradient}
\frac{\partial v_i}{\partial x_j}=\frac{3 Q}{8 \pi r b}
\left[
\begin{smallmatrix}
-\frac{1}{r}\left(1-\frac{z^2}{b^2}\right)  &                                         0 & -\frac{2}{b}\frac{z}{b} \\
                                          0 & \frac{1}{r}\left(1-\frac{z^2}{b^2}\right) &                       0 \\
                                          0 &                                         0 &                       0 \\
\end{smallmatrix}
\right]
\end{eqnarray}
where $z$ is the gap height location between the mold walls, $b$ is half the gap height thickness, and $Q$ is the flow rate.  Orientation results are presented in Figure~\ref{fig:disc_zb_0_4_CI_0_01_FEC} for a gap height of $z/b=4/10$ for isotropic diffusion with $C_I=10^{-2}$ and $\lambda = 1$.  Again, the Hybrid overshoots the actual orientation state, whereas the ORT and the FEC behave in a graphically identical fashion.  This last result further demonstrates the robustness of the FEC approach. Similar tests were performed for gap heights of $z/b=0,1/10,2/10,\ldots,9/10$ and similar conclusions were observed at all gap heights.

\subsection{Results: Computational Time Enhancement}
\label{subsec:results_Time}
An additional goal for any new closure is that of reducing the computational requirements for numerical solutions. Simulations are performed using in-house developed single threaded code using Intel's FORTRAN 90 compiler version 11.1.  Computations are solved on a standard desktop with an Intel i7 processor with 8 GB of Ram. The solution of the ORT has been studied by the investigators for several years, and a reasonably efficient algorithm has been developed. Solutions for the IBOF were made using the FORTRAN 90 code discussed in Jack {\it et al.} \cite{jack:10a}.

Notice from Equations~\eqref{A} and~\eqref{B} that the operations ${\mathbb C}:\left[\cdots\right]$ and ${\mathbb D}:\left[\cdots\right]$ are independent of coordinate frame. As we explained in equation~\eqref{B on N}, in the principal frame there are a considerable number of terms in both ${\mathbb C}$ and ${\mathbb D}$ that are zero that are known prior to any calculations, and thus operations involving $0$ can be avoided in the coding. In addition, computing $DA/Dt$ and $DB/Dt$ in the principle reference frame and then rotating the resulting $3\times3$ tensors into the local reference frame will be more efficient than rotating the $3\times3\times3\times3$ tensors ${\mathbb C}$ and ${\mathbb D}$ into the local reference frame and then computing $DA/Dt$ and $DB/Dt$. All computations of the FEC utilize this characteristic, and thus greatly reduce the computational efforts. In addition, redundant calculations from Equations~\eqref{A} and~\eqref{B} are closely followed and performed only once.  These computations are particularly frequent in the double contractions of the fourth-order tensors with the second-order tensors.

In the first study, computations were performed for the previous closure operations for the ORT and the Hybrid using algorithms similar to implementations discussed in the literature. In studies using an adaptive step size solver, solutions for the IBOF took nearly 10 times that of the ORT, whereas for the fixed step size the two closures required similar computational efforts. To avoid any computational comparisons introduced by an adaptive step size solver, computations were performed using a fixed step-size fourth-order Runge-Kutta (R-K) solver with a very small step size of $\Delta Gt = 10^{-4}$. Computational times are tabulated in Table~\ref{time_table} for both CPU time and normalized time.  Normalized time is defined based off of the often employed Hybrid closure using the standard implementation for the Hybrid closure with the very small step size. The ORT required nearly 770 seconds, a factor of 31 times greater than that of the Hybrid. Conversely, the FEC required only 26 seconds, a slight increase in effort beyond the Hybrid, which required 25 seconds. This is very striking as the Hybrid closure is often selected in research and industrial codes due to its computational efficiency, while recognizing the sacrifice in computational accuracy. This is no longer the case with the FEC as it has the same accuracy of the orthotropic closures while providing computational speeds nearly identical to that of the Hybrid closure.

In the process of developing the FEC algorithm, it was observed that many redundant operations existed in the implementation of the ORT and the Hybrid closures. For existing implementation of the classical closures, no special consideration was given to the ${\mathbb A}:\Gamma$ term, but since the rank four tensor ${\mathbb A}$ is symmetric, equation ~\eqref{B on N} can be used to reduce the number operations of the double contraction to that of a simple rank two tensor operations for both the hybrid closure and the ORT closure implementations. For the ORT, the computational problem can be further simplified by constructing the second-order tensor $DA/Dt$ in the principal frame, and then performing the tensor rotation back into the reference frame.  Thus the costly rotations of the fourth-order tensor ${\mathbb A}$ are avoided. These optimized results for the Hybrid and the ORT are shown in Table~\ref{time_table}, and it is clear that the computational times were greatly reduced.  The optimized Hybrid implementation reduced the computational time to 30\% of the original time, whereas the ORT implementation improved by over an order of magnitude. With these additional computational advances the ORT appears to be a more viable alternative to the Hybrid, but the FEC still has similar computational requirements. It is expected that with further studies, the FEC algorithm could be improved to further reduce its computational times.

\section{Conclusion}
The Fast Exact Closure is a robust, computationally efficient, approach to solve the fiber orientation equations of motion for the orientation tensors.  This unique approach does not require any form of curve fitting based on orientation data obtained from numerical solutions of the full fiber orientation distribution.  The results presented demonstrate that the FEC is as accurate and robust as the existing industrially accepted closures, while enjoying computational speeds equivalent to the industrial form of the hybrid closure.

\section{Acknowledgments}
The authors gratefully acknowledge support from the N.S.F. via grant C.M.M.I. 0727399 and Baylor University through their financial support through their faculty member start-up package.

\section*{Appendix: Justification and Proofs}
By \cite{montgomery-smith:10b}, the Exact Closure is this.  Given $A$, compute the symmetric matrix $B$ by solving
\begin{equation}
\label{A(B)}
A = A(B) = \tfrac12 \int_0^\infty \frac{(B+sI)^{-1} \, ds}{\sqrt{\text{det}(B+s I)}}
\end{equation}
It was shown in \cite{montgomery-smith:10b} that $B$ is unique with this property.  Then compute $\mathbb A$ using the formula
\begin{equation}
\label{bbA}
\mathbb A = \tfrac34 \int_0^\infty \frac{s \, \mathcal S((B+sI)^{-1}\otimes(B+sI)^{-1}) \, ds}{\sqrt{\text{det}(B+s I)}}
\end{equation}
Here $\mathcal S$ represents the symmetrization of a rank 4 tensor, that is, $\mathcal S(\mathbb B)_{ijkl}$ is the average of $\mathbb B_{mnpq}$ over all permutations $(m,n,p,q)$ of $(i,j,k,l)$.

It can be shown that the following two statements are equivalent:
\begin{enumerate}
\item Equation~\eqref{A(B)} holds for all time.
\item Equation~\eqref{A(B)} holds at $t=0$, and equation~\eqref{dadb-C} holds for all time, where
\begin{equation}
\label{C}
\mathbb C = \tfrac34 \int_0^\infty \frac{\mathcal S((B+sI)^{-1}\otimes(B+sI)^{-1}) \, ds}{\sqrt{\text{det}(B+s I)}}
\end{equation}
\end{enumerate}
Furthermore, it can be shown for every symmetric matrix $M$ that
\begin{equation}
\label{BMCM}
\tr (B^{-1}\cdot M) = 2 \tr (\mathbb C:M)
\end{equation}
and hence it can be seen that $\tr (DA/Dt) = 0$ if and only if $\tr (B^{-1}\cdot (DB/Dt)) = 0$, that is, $\tr A$ stays constant if and only if $\det B$ stays constant.

Next, we have
\begin{equation}
\label{invertible}
\text{The linear map on symmetric matrices $M \mapsto \mathbb C:M$ is invertible}
\end{equation}
that is, there exists a rank 4 tensor $\mathbb D$ such that
\begin{equation}
\label{CDM}
\mathbb C:\mathbb D:M = \mathbb D:\mathbb C:M = M \text{ for any symmetric matrix $M$}
\end{equation}
Indeed if we define the six by six matrix
\begin{equation}
\mathcal C = \left[\begin{smallmatrix}
\mathbb C_{1111} & \mathbb C_{1122} & \mathbb C_{1133} &2\mathbb C_{1112} &2\mathbb C_{1113} &2\mathbb C_{1123} \\
\mathbb C_{2211} & \mathbb C_{2222} & \mathbb C_{2233} &2\mathbb C_{2212} &2\mathbb C_{2213} &2\mathbb C_{2223} \\
\mathbb C_{3311} & \mathbb C_{3322} & \mathbb C_{3333} &2\mathbb C_{3312} &2\mathbb C_{3313} &2\mathbb C_{3323} \\
2\mathbb C_{1211} & 2\mathbb C_{1222} & 2\mathbb C_{1233} &4\mathbb C_{1212} &4\mathbb C_{1213} &4\mathbb C_{1223} \\
2\mathbb C_{1311} & 2\mathbb C_{1322} & 2\mathbb C_{1333} &4\mathbb C_{1312} &4\mathbb C_{1313} &4\mathbb C_{1323} \\
2\mathbb C_{2311} & 2\mathbb C_{2322} & 2\mathbb C_{2333} &4\mathbb C_{2312} &4\mathbb C_{2313} &4\mathbb C_{2323}
\end{smallmatrix}\right]
\end{equation}
then $\mathbb D$ can be calculated using the formula
\begin{gather}
\left[\begin{smallmatrix}
\mathbb D_{1111} & \mathbb D_{1122} & \mathbb D_{1133} &\mathbb D_{1112} &\mathbb D_{1113} &\mathbb D_{1123} \\
\mathbb D_{2211} & \mathbb D_{2222} & \mathbb D_{2233} &\mathbb D_{2212} &\mathbb D_{2213} &\mathbb D_{2223} \\
\mathbb D_{3311} & \mathbb D_{3322} & \mathbb D_{3333} &\mathbb D_{3312} &\mathbb D_{3313} &\mathbb D_{3323} \\
\mathbb D_{1211} & \mathbb D_{1222} & \mathbb D_{1233} &\mathbb D_{1212} &\mathbb D_{1213} &\mathbb D_{1223} \\
\mathbb D_{1311} & \mathbb D_{1322} & \mathbb D_{1333} &\mathbb D_{1312} &\mathbb D_{1313} &\mathbb D_{1323} \\
\mathbb D_{2311} & \mathbb D_{2322} & \mathbb D_{2333} &\mathbb D_{2312} &\mathbb D_{2313} &\mathbb D_{2323}
\end{smallmatrix}\right]
= \mathcal C^{-1} \\
\mathbb D_{ijkl} = \mathbb D_{jikl} = \mathbb D_{ijlk}
\end{gather}
In the basis of orthonormal eigenvectors of $B$, since $\mathbb C_{ijkk} = 0$ whenever $i \ne j \ne k$, this reduces to equation~\eqref{D}.

Next, if $B$ is diagonal, then $A$ is diagonal with entries
\begin{equation}
\label{A long}
\begin{split}
a_{1} = \tfrac12 \int_0^\infty \frac{ds}{(b_1+s)^{3/2}\sqrt{b_2+s}\sqrt{b_3+s}} \\
a_{2} = \tfrac12 \int_0^\infty \frac{ds}{\sqrt{b_1+s}(b_2+s)^{3/2}\sqrt{b_3+s}} \\
a_{3} = \tfrac12 \int_0^\infty \frac{ds}{\sqrt{b_1+s}\sqrt{b_2+s}(b_3+t)^{3/2}}
\end{split}
\end{equation}
and
\begin{equation}
\label{C long}
\begin{split}
\mathbb C_{1111} = \tfrac34 \int_0^\infty \frac{ds}{(b_1+s)^{5/2}\sqrt{b_2+s}\sqrt{b_3+s}} & \qquad \quad
\mathbb C_{1122} = \tfrac14 \int_0^\infty \frac{ds}{(b_1+s)^{3/2}(b_2+s)^{3/2}\sqrt{b_3+s}} \\
\mathbb C_{2222} = \tfrac34 \int_0^\infty \frac{ds}{\sqrt{b_1+s}(b_2+s)^{5/2}\sqrt{b_3+s}}  & \qquad \quad
\mathbb C_{2233} = \tfrac14 \int_0^\infty \frac{ds}{\sqrt{b_1+s}(b_2+s)^{3/2}(b_3+s)^{3/2}} \\
\mathbb C_{3333} = \tfrac34 \int_0^\infty \frac{ds}{\sqrt{b_1+s}\sqrt{b_2+s}(b_3+t)^{5/2}} & \qquad \quad
\mathbb C_{1133} = \tfrac14 \int_0^\infty \frac{ds}{(b_1+s)^{3/2}\sqrt{b_2+s}(b_3+s)^{3/2}}
\end{split}
\end{equation}
all the other coefficients of $\mathbb C$ being zero.  Furthermore
\begin{equation}
\label{C:I}
\mathbb C : I = \tfrac12 B^{-1}
\end{equation}
Equations~\eqref{C quick} now follow by an easy calculation.  Equations~\eqref{C quick b1=b2+epsilon} and~\eqref{C quick b1=b2=b3+epsilon} are obtained by expanding the fourth equation of~\eqref{C long} using Taylor's series, where
\begin{equation}
\label{In def}
\mathcal I_n = \int_0^\infty \frac{ds}{(b_0+s)^n\sqrt{b_3+s}}
\end{equation}
The proofs of various details now follow.

\proof{Proof of equations~\bfeqref{dadb-C} and~\bfeqref{C}:}
Write $\dot A$ and $\dot B$ for $\frac{DA}{Dt}$ and $\frac{DB}{Dt}$ respectively.  Use the formulae
\begin{equation}
\label{proof_of_dadb-C}
\frac {D}{Dt} B^{-1} = - B^{-1} \cdot \dot B \cdot B^{-1}
\quad\text{and}\quad
\frac {D}{Dt} \det B = \tr (B^{-1} \cdot \dot B) \det B
\end{equation}
to obtain
\begin{equation}
\begin{split}
\dot A & =
- \tfrac12 \int_0^\infty \frac{(B+sI)^{-1} \cdot \dot B \cdot (B+sI)^{-1} \, ds}{\sqrt{\text{det}(B+s I)}}
- \tfrac14 \int_0^\infty \frac{[(B+sI)^{-1}:\dot B] \, (B+sI)^{-1} \, ds}{\sqrt{\text{det}(B+s I)}} \\
& =
-\mathbb C:\dot B
\end{split}
\end{equation}
since for any symmetric matrix $K$ we have
\begin{equation}
\label{S_of_dyad}
\mathcal S(K\otimes K):\dot B = \tfrac13 K\cdot\dot B \cdot K + \tfrac23 (K:\dot B) K
\end{equation}

\proof{Proof of equation~\bfeqref{CBM}:}  For any invertible symmetric matrix $K$
\begin{equation}
\label{proof_of_CMB_symmetrix_matrix}
\mathcal S(K^{-1}K^{-1}):(K\cdot M)
=
\mathcal S(K^{-1}K^{-1}):(M^T\cdot K)
=
\tfrac13 ((\tr M) K^{-1} + M\cdot K^{-1} + K^{-1}\cdot M^T)
\end{equation}
Setting $K = B+sI$, we multiply both sides by $3/(4\sqrt{\det(B+sI)})$, and integrate with respect to $s$ from zero to infinity, to obtain equation~\eqref{CBM}.


\proof{Proof of equations~\bfeqref{In} from \bfeqref{In def}:}  To compute $\mathcal I_1$, use the formulae
\begin{gather}
\frac{d}{ds}\cos^{-1}\left(\sqrt{\frac{b_3+s}{b_0+s}}\right) = -\frac{\sqrt{b_0-b_3}}{2(b_0+s)\sqrt{b_3+s}} \\
\frac{d}{ds}\cosh^{-1}\left(\sqrt{\frac{b_3+s}{b_0+s}}\right) = -\frac{\sqrt{b_3-b_0}}{2(b_0+s)\sqrt{b_3+s}}
\end{gather}
Next, integrating by parts, we obtain
\begin{equation}
\label{In expansion}
\mathcal I_n = - \frac{\sqrt{b_3}}{2 b_0^n} + \frac n2 \int_0^\infty \frac{\sqrt{b_3+s} \, ds}{(b_0+s)^{n+1}}
\end{equation}
and simple algebra gives
\begin{equation}
\label{In expansion2}
\int_0^\infty \frac{\sqrt{b_3+s} \, ds}{(b_0+s)^{n+1}} = \mathcal I_n + (b_3-b_0) \mathcal I_{n+1}
\end{equation}

\proof{Proof of equation~\bfeqref{BMCM}:}  For any positive definite matrix $X$, if
\begin{equation}
A(X) = \tfrac12 \int_0^\infty \frac{(X+sI)^{-1} \, ds}{\sqrt{\text{det}(X+s I)}}
\end{equation}
then
\begin{equation}
\tr (A(X)) = \tfrac12 \int_0^\infty \frac{\tr ((X+sI)^{-1}) \, ds}{\sqrt{\text{det}(X+s I)}}
= - \int_0^\infty \frac{d}{ds}\left(\frac1{\sqrt{\text{det}(X+s I)}}\right) \, ds
= \frac{1}{\sqrt{\det X}}
\end{equation}
If $\tr (B^{-1}\cdot M) = \alpha$, then (remembering that $\det B = 1$) we have $\det(B+\epsilon M) = 1 + \epsilon \alpha + O(\epsilon^2)$ as $\epsilon\to0$.  Hence $1 - \tfrac12\epsilon \alpha + O(\epsilon^2) = \tr (A(B+\epsilon M)) = 1 - \epsilon \tr (\mathbb C:M) + O(\epsilon^2)$.  Therefore $\tr (\mathbb C:M) = \tfrac12 \alpha$.

\proof{Proof of equation~\bfeqref{invertible}:}
This follows because
\begin{equation}
\label{C pos def}
\text{$M$ is a symmetric non-zero matrix} \Rightarrow M:\mathbb C:M > 0
\end{equation}
and hence $M\ne0 \Rightarrow \mathbb C:M \ne 0$.

To see this, suppose that $K$ is a positive definite three by three matrix, and let $k_1$, $k_2$ and $k_3$ be its eigenvalues.  Then in the basis of corresponding orthonormal eigenvalues of $K$, we have that for any non-zero symmetric $M$
\begin{equation}
M:\mathcal S(K\otimes K):M = \tfrac13\left(\sum_{i=1}^3 k_i M_{ii}\right)^2 + \tfrac23\sum_{i,j=1}^3 k_i k_j M_{ij}^2 > 0
\end{equation}
Apply this to $K=(B+sI)^{-1}$, multiply by $(\det(B+sI))^{-1/2}$, and then integrate over $s$ to obtain $M:\mathbb C:M > 0$.

\proof{Proof of equation~\bfeqref{C:I}:}  Without loss of generality $B$ is diagonal.  Hence we need to prove statements such as $\mathbb C_{1111} + \mathbb C_{1122} + \mathbb C_{1133} = \tfrac12 b_1^{-1}$ when $\mathbb C$ satisfies equation~\eqref{C long}.  But
\begin{equation}
\begin{split}
\mathbb C_{1111} + \mathbb C_{1122} + \mathbb C_{1133}
&=
- \tfrac12 \int_0^\infty \frac{d}{ds} \left(\frac1{(b_1+s)^{3/2}\sqrt{b_2+s}\sqrt{b_3+s}}\right)\,ds \\
&=
\tfrac12 b_1^{-3/2} b_2^{-1/2} b_3^{-1/2}
\end{split}
\end{equation}
The result follows since $b_1 b_2 b_3 = 1$.

\proof{Proof of equation~\eqref{FEC-A-generic-rsc}:}
From equation~\eqref{dadb-C}, we see that the RSC version of equation~\eqref{FEC-A-generic} is equation~\eqref{FEC-A-generic-rsc} and
\begin{equation}
\frac{DB}{Dt} = G(A,B) - (1-\kappa)\mathbb D:\mathbb M:F(A,B)
\end{equation}
Since $\mathbb D:F(A,B) = G(A,B)$, it follows that all we need to show is $\mathbb D:\mathbb M:F(A,B) = \mathbb M:\mathbb D:F(A,B)$.  This is easily seen by working in the basis of orthonormal eigenvectors of $B$, noticing that then $\mathbb M:N$ is simply the diagonal part of $N$, and applying equation~\eqref{B on N}.

\proof{Proof of Theorem~\ref{thm-ft}:}
It follows from $\det B = 1$ that the only way that the solutions can become non-physical is if $B$ `blows up,' that is, if one or more of the eigenvalues of $B$ become infinite in finite time.  (Also, \cite[Theorem~1.4]{chicone:06} can be used to show that the finiteness of the eigenvalues of $B$ imply the differential equations have a unique solution.)

Substituting $M=I$ into equation~\eqref{CBM}, we obtain $\mathbb C:B = \tfrac32 A$, that is,
\begin{equation}
\label{DAB}
\mathbb D:A = \tfrac23 B
\end{equation}
Take the trace of equation~\eqref{B} and use equation~\eqref{DAB}, to obtain
\begin{equation}
\frac{D}{D t} \tr  B \le c(\|\Omega\| + \|\Gamma\| + D_r) (\tr B) - 2 D_r (I:\mathbb D:I)
\end{equation}
for some universal constant $c>0$.  Here $\|\cdot\|$ denotes the spectral norm of a matrix, and we have used the inequality $\tr(X\cdot Y) \le \|X\| \tr Y$ whenever $Y$ is positive definite.  By equation~\eqref{C pos def}, we have $I:\mathbb D:I = M:\mathbb C:M \ge 0$, where $M = \mathbb D:I$, and hence
\begin{equation}
\frac{D}{D t} \tr  B \le c(\|\Omega\| + \|\Gamma\| + D_r) (\tr B)
\end{equation}
Now we can apply Gronwall's inequality \cite[Chapter~2.1.1]{chicone:06} (in Lagrangian coordinates) to obtain
\begin{equation}
\tr B \le (\tr B_0) e^{c t L}
\end{equation}
where $L$ is an upper bound for $\|\Omega\| + \|\Gamma\| + D_r$, and $B_0$ is the value of $B$ at $t=0$.  Therefore $\tr B$ remains finite, and since $B$ is positive definite, no eigenvalue of $B$ blows up to infinity in finite time.

\proof{Proof of Theorem~\ref{thm-Koch}:}  Note that the positive definiteness of $D_r$, and the boundedness of $D_r$ and $1/\|D_r^{-1}\|$ guarantee that the ratio of $\tr B$ and $D_r:B$ is bounded from above and below.  From equation~\eqref{B-A-Koch}, and using equation~\eqref{DAB}
\begin{equation}
\frac{D}{D t} (D_r:B) \le {}
 c\left(\|\Omega\| + \|\Gamma\| + \|D_r\| + \left\|\frac{D(D_r)}{Dt}\right\|\right) (D_r:B)
 - 2 (D_r:\mathbb D:D_r)
\end{equation}
The rest of the proof proceeds by a similar argument as above.

\section{References}

\bibliographystyle{elsarticle-num}
\bibliography{info_10}

\gdef\thefigure{\arabic{figure}}
\gdef\thetable{\arabic{table}}
\pagebreak \clearpage

\begin{table}[ht]
\begin{center}
\begin{tabular}{ccccccccccc}\hline\hline \vspace{-10pt}         \\
Flow \# & $v_1$            & $v_2$   & $v_3$     & $C_I$     & $\lambda$ & $\overline{\varepsilon}_{\mbox{\tiny FEC}}$ & $\overline{\varepsilon}_{\mbox{\tiny ORT}}$ & $\overline{\varepsilon}_{\mbox{\tiny IBOF}}$ & $\overline{\varepsilon}_{\mbox{\tiny Hybrid}}$ \\ \hline\hline
1   &  $Gx_1$           & $Gx_2$  & $-2Gx_3$  & $10^{-3}$ & 1    & 1    &  2.06  &  1.28  &  76.2\\
2   &  $2Gx_1$          & $-Gx_2$ & $-Gx_3$   & $10^{-3}$ & 1    & 1.03 &  1.05  &  1     &  25.8\\
3a  &  $G x_3$          & $0$     & $0$       & $10^{-3}$ & 0.99 & 1    &  1.02  &  1.02  &  3.69\\
3b  &  $G x_3$          & $0$     & $0$       & $10^{-3}$ & 1    & 1    &  1.02  &  1.01  &  3.28\\
4   &  $-Gx_1+10Gx_2$   & $-Gx_2$ & $2Gx_3$   & $10^{-3}$ & 1    & 1    &  2.25  &  1.36  &  12.9\\
5   &  $-Gx_1+Gx_2$     & $-Gx_2$ & $2Gx_3$   & $10^{-3}$ & 1    & 1.02 &  1     &  1.23  &  22.6\\
6   &  $Gx_1+2Gx_3$     & $Gx_2$  & $-2Gx_3$  & $10^{-2}$ & 1    & 1    &  1.01  &  1.08  &  3.57\\
7   & $Gx_1+2.75Gx_3$   & $Gx_2$  & $-2G x_3$ & $10^{-2}$ & 1    & 1    &  1.02  &  1.05  &  2.98\\
8   &  $Gx_1+1.25Gx_3$  & $Gx_2$  & $-2G x_3$ & $10^{-2}$ & 1    & 1.02 &  1     &  1.12  &  3.85\\
9   & $-Gx_1+10G x_3$   & $Gx_2$  & $0$       & $10^{-2}$ & 1    & 1.03 &  1     &  1.03  &  1.65\\
10  &  $-Gx_1+G x_3$    & $Gx_2$  & $0$       & $10^{-2}$ & 1    & 1.01 &  1     &  1.04  &  2.29\\
11  &  $2Gx_1+3Gx_3$    & $-Gx_2$ & $-Gx_3$   & $10^{-2}$ & 1    & 1.04 &  1.04  &  1     &  2.52\\
12  &  $-Gx_1+3.75Gx_2$ & $Gx_2$  & $2Gx_3$   & $10^{-2}$ & 1    & 1    &  1.03  &  1.06  &  2.03\\
13  &  $-Gx_1+1.5G x_2$ & $-Gx_2$ & $2Gx_3$   & $10^{-2}$ & 1    & 1.00 &  1     &  1.01  &  2.34\\
14a &  $Gx_3$           & $0$     & $0$       & $10^{-2}$ & 0.99 & 1    &  1.00  &  1.03  &  4.14\\
14b &  $Gx_3$           & $0$     & $0$       & $10^{-2}$ & 1    & 1    &  1.00  &  1.02  &  3.90
\\
\hline\hline
\end{tabular}
\caption{Flows used, and the resulting error computation in computing the second-order orientation tensor $A$.}
\label{norm_error_table}
\end{center}
\end{table}

\begin{table}[h]  \begin{center}
\begin{tabular}{ccc}\hline\hline \vspace{-10pt}         \\
Closure & CPU Time & Normalized Time \\
\hline
Hybrid - Original   &   25    &   1  \\
Hybrid - Optimized  &   6.9   &  0.3 \\
ORT - Original	    &   770   &   31 \\
ORT - Optimized	    &    21   & 0.8  \\
FEC	                &    26   & 1.0  \\
\hline\hline
\end{tabular}
\caption{Normalized Computational Times}
\label{time_table}
\end{center}
\end{table}

\pagebreak \clearpage

\pagebreak \clearpage

\pagebreak \clearpage


\end{document}